\documentclass[journal,comsoc]{IEEEtran}

\usepackage{amsmath}
\usepackage{amsfonts,amssymb}
\usepackage{multirow}
\usepackage[dvips]{graphicx}

\def\FF{\mathbb{F}}

\def\Label#1{\label{#1}\ [\ \text{#1}\ ]\ }
\def\Label{\label}

\begin{document}
\title{Secure physical layer network coding versus secure network coding
\thanks{The material in this paper was presented in part at the 2018 IEEE Information Theory Workshop (ITW), Guangzhou, China, 25-29 November 2018. \cite{ITW2018}}}

\author{Masahito~Hayashi,~\IEEEmembership{Fellow,~IEEE} 
\thanks{Masahito Hayashi is with
Graduate School of Mathematics, Nagoya University, Nagoya, Japan,
Shenzhen Institute for Quantum Science and Engineering, Southern University of Science and Technology, and
Centre for Quantum Technologies, National University of Singapore, Singapore.
e-mail: masahito@math.nagoya-u.ac.jp}
\thanks{Manuscript submitted 1st December 2018; revised xxx, 2018.}}

\maketitle

\begin{abstract}
Secure network coding realizes the secrecy of the message when the message is transmitted via 
noiseless network
and a part of edges or a part of intermediate nodes are eavesdropped.
In this framework, if the channels of the network has noise,
we apply the error correction to noisy channel before applying the secure network coding.
In contrast, 
secure physical layer network coding
is a method to securely transmit a message by a combination of coding operation on nodes
when the network is given as a set of noisy channels.
In this paper, we give several examples of network, in which,
secure physical layer network coding 
has advantage over 
secure network coding.
\end{abstract}


\section{Introduction}
Secure network coding is a method to securely transmit the message via 
noiseless network when a part of edges or a part of intermediate nodes are eavesdropped \cite{Cai2002,CY,YN,CG,RSS,FMSS}.
Since the real channel has noise, we apply the error correction to the real channel.
Then, we apply secure network coding to the noiseless channel realized by error correction.
That is, in the above scenario, 
we separately apply the error correction and secure network coding.
Therefore, there is a possibility that
we have an advantage by jointly applying the error correction and secure network coding.
This idea is called physical network coding \cite{ZLL,NG3,PY}.
That is, to consider this improvement for the security, 
we discuss the secure version of physical layer network coding, i.e.,
secure physical layer network coding, which 
is a method to securely transmit a message by a combination of coding operation on nodes
when the network is given as a set of noisy channels.
There are two kinds of codes in 
secure physical layer network coding.
Once we have secure network coding, 
we can attach physical layer network coding.
This method can be considered as a simple combination of 
secure network coding and physical layer network coding.
The other type of codes in secure physical layer network coding
are codes that cannot be made by such a simple combination.
Unfortunately, there are almost no studies for secure physical layer network coding of the latter type.
That is, existing studies addressed only 
secure computation-and-forward, which is a method to securely transmit 
the modulo sum of two input message when noisy multiple access channel is given \cite{Ren,He1,He2,Vatedka,Zewail,arXiv}.
The motivation of these studies is the realization of secure two way-relay channel with untrusted relay.
To seek the further possibility of secure physical layer network coding, 
we need to find more examples of concrete coding schemes of 
secure physical layer network coding.

In fact, secure network code mainly focuses on the secrecy for the attack to channels.
Some typical secure network codes cannot realize the secrecy
when one of intermediate nodes is eavesdropped.
In contrast, secure physical layer network coding is advantageous for attacks on intermediate nodes.
In this paper, we give two examples of network, in which,
secure physical layer network coding realizes a performance that cannot be realized by
secure network coding.
One is butterfly network \cite{ACLY} and the other is a network with three source nodes.

The remaining parts of this paper are organized as follows.
Section \ref{S2} reviews the results of secure computation-and-forward,
which is a typical example of secure physical layer network coding.
 Section \ref{S3} discusses secure communication over butterfly network
by using secure physical layer network coding.
 Section \ref{S4} addresses secure communication over a network with three source nodes
by using secure physical layer network coding.

\section{Secure computation-and-forward}\label{S2}
\begin{figure}[b]
\begin{center}
\includegraphics[scale=0.6]{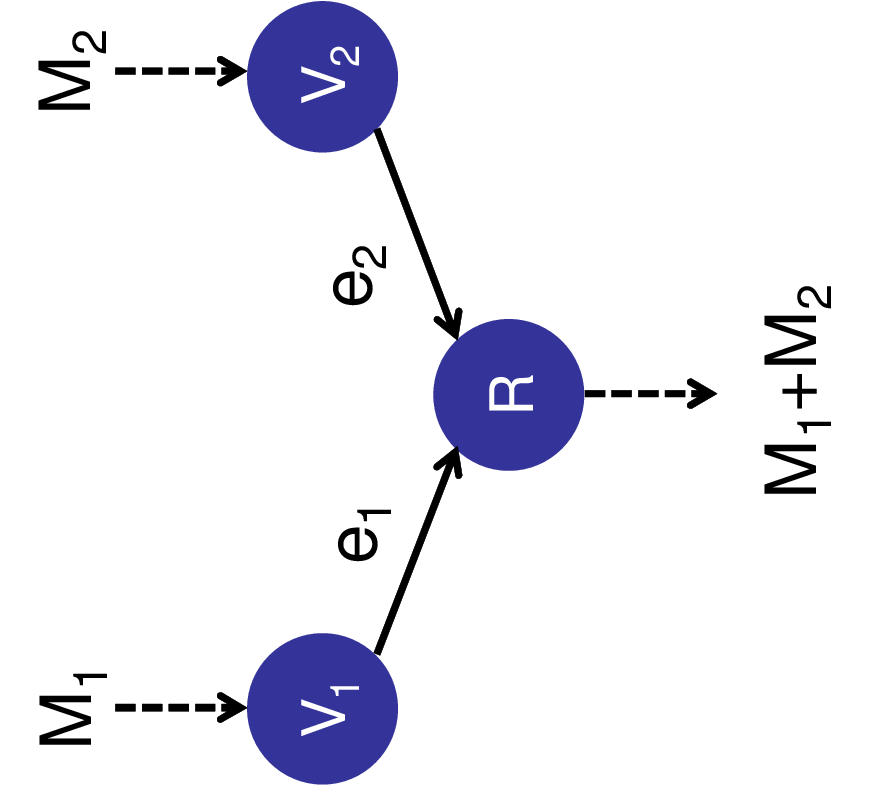}
  \end{center}
\caption{Computation-and-forward.}
\Label{CaF}
\end{figure}   
First, we review secure computation-and-forward.
We consider secure computation-and-forward in a typical setting.
Consider two senders $V_1$, $V_2$ and one receiver $R$.
Assume that the sender $V_i$ has message $M_i\in \FF_q$,
and the receiver $R$ 
are linked by a (noisy) multiple access channel that has two inputs from the two senders $V_1$ and $V_2$.
Computation-and-forward is the task for the receiver $R$ to obtain the modulo sum $M_1+M_2$ via the (noisy) 
multiple access channel as Fig. \ref{CaF}.

To discuss computation-and-forward, 
many papers focused on a multiple access Gaussian channel.
When the sender $V_i$ sends the complex-valued signal $X_i$ for $i=1,2$,
the receiver $R$ receives the complex-valued signal $Y$ as 
\begin{align}
Y= h_1 X_1 + h_2 X_2 +Z,
\label{MAC}
\end{align}
where $h_1, h_2 \in \mathbb{C}$ are the channel fading coefficients,
and $Z $ is a Gaussian complex random variable
with average 0 and variance 1.
In the following of this section,
we assume multiple use of the above multiple access Gaussian channel.

Using lattice codes,
the papers \cite{NG1,NG2,NCNC} derived an achievable rate under the energy constraint, which is called the computation rate.
Also, 
based on the BPSK scheme, in which $X_i$ is coded to $(-1)^{A_i}$ with $A_{i}\in \FF_2$,
the paper \cite{Ullah} derived an achievable rate $I(Y; A_1+A_2)_{\rm Eq. \eqref{MAC}}$,
where the mutual information is given with
the independent and uniform random numbers $A_1$ and $A_2$.
In this paper, we choose the base of logarithm to be $e$.
Then, 
the papers \cite{Sula,Takabe} proposed to use LDPC codes 
(regular LDPC codes and spatial coupling LDPC codes)
with the BPSK scheme.
The method proposed by \cite{Sula,Takabe} can be efficiently realized,
and realizes a rate close to $I(Y; A_1+A_2)_{\rm Eq. \eqref{MAC}}$.

When we additionally impose the secrecy for each message to the receiver $R$,
this task is called secure computation-and-forward.
In this case, it is required that the receiver $R$ obtains the modulo sum $M_1+M_2$,
but the variable in $R$'s hand is independent of $M_1$ and $M_2$.
The papers \cite{Ren,He1,He2,Vatedka,Zewail} proposed a code for this task by using lattice code.
Using an arbitrary algebraic code for computation-and-forward given in \cite{Sula,Takabe},
the paper \cite{arXiv} proposed an efficiently realizable code.
The paper \cite{arXiv} also derived an upper bound for the leaked information of 
the constructed finite length code.
Also, the paper \cite{arXiv} showed that the rate 
$2 I(Y;A_1+A_2)_{\rm Eq. \eqref{MAC}} -I(Y; A_1,A_2  )_{\rm Eq. \eqref{MAC}}$
is achievable in the BPSK scheme \cite[(29)]{arXiv},
where the mutual information is given with
the independent and uniform random numbers $A_1$ and $A_2$.
That is, when the channel \eqref{MAC} is prepared and the receiver colludes with no sender,
secure computation-and-forward guarantees no information leakage of each message to the receiver
 while the receiver can recover the sum $M_1+M_2$.
In fact, all these papers \cite{Ren,He1,He2,Vatedka,Zewail} for secure computation-and-forward
addressed only the case when the number of senders is 2.
The paper \cite{Preparation} will address 
secure computation-and-forward 
when the number of senders is more than 2.

Unfortunately, we have no good application for 
secure computation-and-forward 
except for secure two way-relay channel with untrusted relay.
The remaining part of this paper discusses its further application.

\begin{figure}[t]
\begin{center}
\includegraphics[scale=0.7]{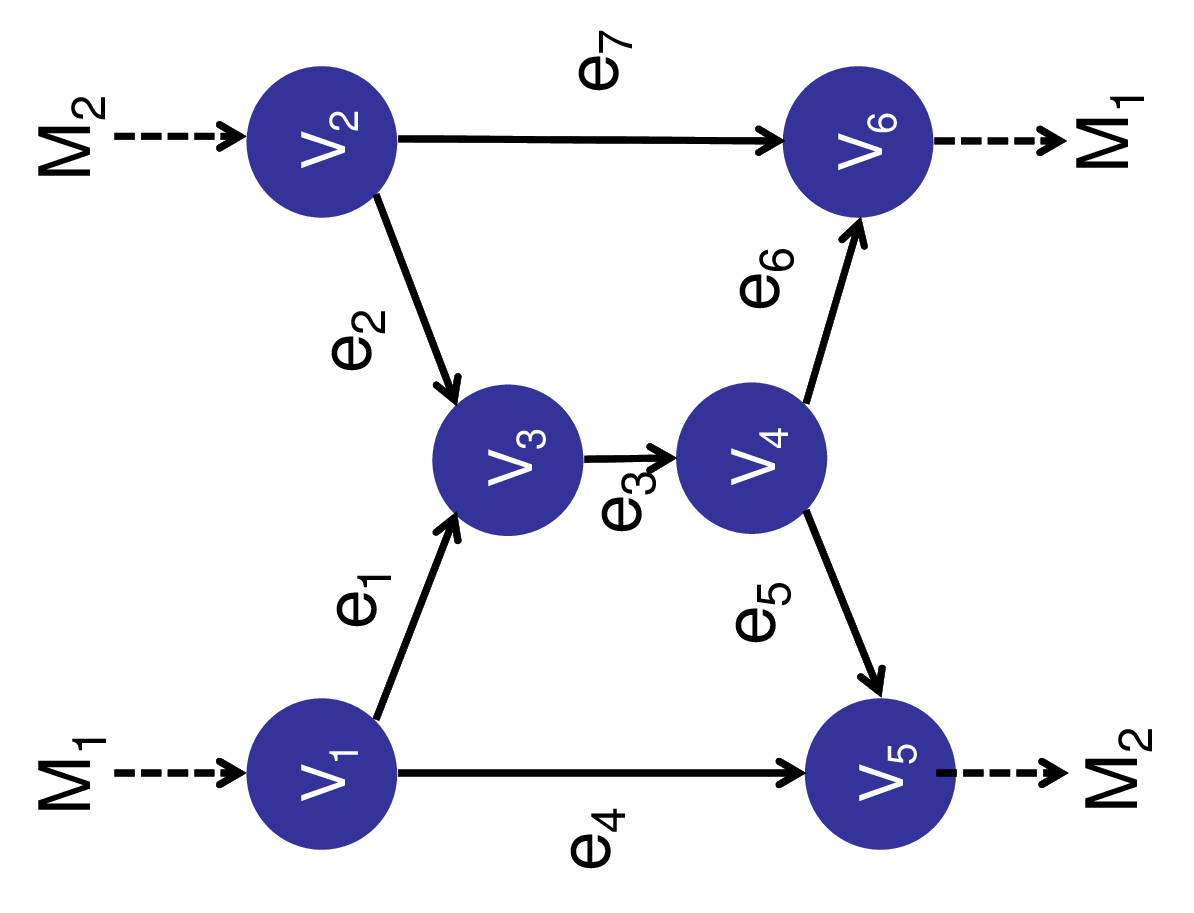}
  \end{center}
\caption{Butterfly network coding.}
\Label{F8}
\end{figure}   

\section{Butterfly network}\label{S3}
\subsection{conventional protocol}\label{S31}
We consider 
butterfly network coding \cite{ACLY}, which is a coding method that efficiently transmits the information 
in the crossing way as Fig. \ref{F8}.
The aim of this network is 
the following two tasks. 
The transmission of the message $M_1$ from $V_1$ to $V_6$
and the transmission of the message $M_2$ from $V_2$ to $V_5$.
When each channel can transmit only one element of $\FF_q$,
the channel $e_3$ from $V_3$ to $V_4$ is the bottleneck.
In this network, only the node $V_3$ has the choice
because other node has no other choice except for transmitting the received information.
To resolve this bottleneck, 
the node $V_3$ transmits the modulo sum to the node $V_4$ via channel $e_3$.
Then, the sink node $V_5$ can recover the message $M_2$ from the received information 
$M_1$ and $M_1+M_2$.
Similarly, the other sink node $V_6$ can recover the message $M_1$ from the received information 
$M_2$ and $M_1+M_2$.

\subsection{Secure network coding}\label{S32}
However, under this network code, 
the node $V_3$ obtains both messages $M_1$ and $M_2$.
The sink node $V_5$ obtains the intended message $M_1$,
and the other sink node $V_6$ has the same problem.
Now, we consider impose the secrecy for the attack to one of intermediate nodes.
That is, the information of all intermediate nodes needs to be independent of $M_1$ and $M_2$,
and the information of source node $V_5$ ($V_6$) needs to be independent of the unintended message $M_1$ ($M_2$).
This kind of secrecy can be realized by employing a shared secret number $L$ between $V_1$ and $V_2$ 
when messages $M_1$ and $M_2$ are elements of $\FF_q$ and $q$ is not a power of $2$ 
in the following way \cite[Figure 2]{OKH}.
When the information transmitted $Z_i$ on the edge $e_i$ is given as
\begin{align*}
Z_1&=2M_1+L, ~Z_4=-(M_1+L),\\
Z_2&=2M_2+L, ~Z_7=-(M_2+L), \\
Z_3&= Z_1+Z_2= 2M_1+2M_2+2L, \\
Z_5 &=Z_6=Z_3/2,\\
\hat{M}_2&= Z_5+Z_4= M_2 ,~
\hat{M}_1= Z_6+ Z_7= M_1 ,
\end{align*}
where $\hat{M}_2$ ($\hat{M}_1$) is the recovered message by $V_5$ ($V_6$).
Any intermediate edge and any intermediate node obtain no information for the messages $M_1$ and $M_2$.
Also, the sink node $V_5$ ($V_6$) obtains no information for the message $M_1$ ($M_2$) while it obtains the message $M_2$ ($M_1$).
Hence, this code guarantees the following types of security;
(B1) When the eavesdropper attacks only one of edges, 
she obtains no information for  the messages $M_1$ and $M_2$.
(B2) When no node colludes with another node,
each node obtains no information for the unintended messages.

When $q \ge 4$ is a power of $2$,
the above code can be modified as follows.
We choose an element $e\in \FF_q$ such that $e^2+e\neq 0$, i.e., 
$e\neq 1,0$.
Then, we arrange our code as 
\begin{align*}
Z_1&=(1+e)M_1+L, ~Z_4=-(M_1+L), \\
Z_2& =(1+e)M_2+e L, ~Z_7=-(M_2+L), \\
Z_3&= Z_1+Z_2= (1+e)(M_1+M_2+L) ,\\
Z_5&=Z_6= Z_3/(1+e) ,\\
\hat{M}_2&= Z_5+Z_4= M_2 ,~
\hat{M}_1= Z_6+Z_7= M_1 .
\end{align*}
This modification realizes the required security 
in this case.

\subsection{Secure physical layer network coding}\label{S33}
But, if there is no  shared secret number between $V_1$ and $V_2$, 
it is not so easy to realize this kind of secrecy for the butterfly network
under the framework of secure network coding.
Now, we assume the assumption; (B3) The pairs $(e_1,e_2)$, $(e_4,e_5)$, and $(e_6,e_7)$ are given as multiple access Gaussian channels like \eqref{MAC}.
Only the channel $e_3$ is a single input Gaussian channel.
In this case, 
in the multiple access Gaussian channel $(e_1,e_2)$ at $V_3$,
we employ secure computation-and-forward so that the node 
$V_3$ obtains the information $M_1+M_2$.
Then, the node $V_3$ forwards the obtained information to the node $V_4$,
and the node $V_4$ receives the information $M_4:=M_1+M_2$.
In the multiple access Gaussian channel $(e_4,e_5)$ at $V_5$,
we again employ secure computation-and-forward so that the node 
$V_5$ obtains the information $M_4-M_1=M_2$.
In the same way,
the node $V_6$ obtains the information $M_4-M_2=M_1$.
That is, this code guarantees the following types of security;
(B4) When no node colludes with another node and (B3) is satisfied,
each node obtains no information for the unintended messages.

As another kind of secure physical layer network coding,
we attach the computation-and-forward 
to the communications to nodes $V_3$, $V_5$, and $V_6$ in the protocol with $q=4$ 
given in Section \ref{S32}.
In this protocol, an element of $\FF_4$ is regarded as a vector on $\FF_2$. 
While this protocol saves the time, it still requires the secure shared randomness $L$.
This protocol can be regarded as a simple combination of 
secure network coding and physical layer network coding.

\subsection{Comparison}\label{S34}
To implement these protocols as wireless communication network,
we compare the transmission speeds of 
the protocols given in Sections \ref{S32} and \ref{S33}
when each edge is given as the BPSK scheme
of a two-input Gaussian channel as \eqref{MAC}
or a single-input Gaussian channel
\begin{align}
Y= h X+Z,
\label{C2}
\end{align}
where
$h \in \mathbb{C}$ are the channel fading coefficients,
$Z $ is a Gaussian complex random variable with average 0 and variance 1,
and $X$ is coded as $(-1)^A$ with $A\in \FF_2$.
In this comparison, for simplicity, we assume that $h_1=h_2=h$.
Here, we assume that $T$ is the time period to transmitting one Gaussian signal in each edge. 
Additionally, we assume that ideal codes are available as follows.
The mutual information rate $I(Y;A)_{\rm Eq. \eqref{C2}}$ is available in the channel \eqref{C2}, 
the rate 
$ I(Y;A_1+A_2)_{\rm Eq. \eqref{MAC}} $
 is available for computation-and-forward in the channel \eqref{MAC},
 and
the rate 
$2 I(Y;A_1+A_2)_{\rm Eq. \eqref{MAC}} -I(Y; A_1,A_2  )_{\rm Eq. \eqref{MAC}}$
 is available for secure computation-and-forward in the channel \eqref{MAC}.
Notice that the relation $ I(Y;A_2|A_1 )_{\rm Eq. \eqref{MAC}} = I(Y;A_1|A_2 )_{\rm Eq. \eqref{MAC}}$ holds in this case.
Also, the mutual information rate pair 
$(I(Y;A_1A_2)_{\rm Eq. \eqref{MAC}}/2,I(Y;A_1A_2)_{\rm Eq. \eqref{MAC}}/2)$ 
is available in the MAC channel \eqref{MAC}
when both senders intend to send their own message to the receiver.
In the above discussion, the random variables $A_1$, $A_2$, and $A$ are subject to the uniform distribution independently.

Secure network coding protocol given in Section \ref{S32} needs to avoid a crossed line 
when we do not use multiple access Gaussian channel. 
Hence, its whole network needs four time slots at least as Table \ref{T1}.
Therefore, to transmit message with size $e^{R}$, the required time in this case is
calculated to be $ \frac{4RT}{I(Y;A)_{\rm Eq. \eqref{C2}}}$.

\begin{table}[htpb]
  \caption{Secure network coding without multiple access Gaussian channel}
\label{T1}
\begin{center}
{
\renewcommand\arraystretch{1.7}
  \begin{tabular}{|c|c|c|c|c|} 
\hline
Time slot &Time 1 & Time 2 & Time 3 & Time 4  \\
\hline
Channel &$e_1$, $e_4$ &$e_2$, $e_7$ & $e_3$ & $e_5$, $e_6$ \\
\hline
  \end{tabular}}
\end{center}
\end{table}

When we use multiple access Gaussian channel,
secure network coding protocol given in Section \ref{S32}
can be realized with three time slots as Table \ref{T12}.
In this case, to transmit message with size $e^{R}$, the required time is
calculated to be 
$ \frac{2RT}{I(Y;A)_{\rm Eq. \eqref{C2}}}+ \frac{2RT}{I(Y;A_1,A_2)_{\rm Eq. \eqref{MAC}}}$.
Although we can design the whole process as Table \ref{T2},
this design requires the time length  
$ \frac{RT}{I(Y;A)_{\rm Eq. \eqref{C2}}}+ \frac{4RT}{I(Y;A_1,A_2)_{\rm Eq. \eqref{MAC}}}$,
which is larger than 
$ \frac{2RT}{I(Y;A)_{\rm Eq. \eqref{C2}}}+ \frac{2RT}{I(Y;A_1,A_2)_{\rm Eq. \eqref{MAC}}}$
because $\frac{I(Y;A_1,A_2)_{\rm Eq. \eqref{MAC}}}{2} 
\le I(Y;A)_{\rm Eq. \eqref{C2}}$.

\begin{table}[htpb]
  \caption{Secure network coding with multiple access Gaussian channel}
\label{T12}
\begin{center}
{
\renewcommand\arraystretch{1.7}
  \begin{tabular}{|c|c|c|c|} 
\hline
Time slot &Time 1 & Time 2 & Time 3  \\
\hline
Channel &$(e_1, e_2)$ &$e_3$, $e_4$, $e_7$ & $e_5$, $e_6$ \\
\hline
  \end{tabular}}
\end{center}
$(e_i, e_j)$ expresses a multiple access Gaussian channel of the joint transmission on the edges $e_i$ and $e_j$.
\end{table}

Secure physical layer network coding protocol given in the 1st paragraph of Section \ref{S33} 
can be realized only with three time slots as in Table \ref{T2},
where the pairs $(e_1, e_2)$, $(e_4,e_5)$, and $(e_6,e_7)$
are realized by secure computation-and-forward
based on the Gaussian MAC channel \eqref{MAC}.
Therefore, to transmit message with size $e^{R}$, the required time in this case is
calculated to be 
$ \frac{2 RT}{2 I(Y;A_1+A_2)_{\rm Eq. \eqref{MAC}} -I(Y; A_1,A_2  )_{\rm Eq. \eqref{MAC}}}+\frac{RT}{I(Y;A)_{\rm Eq. \eqref{C2}}}$.

Secure physical layer network coding protocol given in the 2nd paragraph of Section \ref{S33} 
also can be realized only with three time slots as in Table \ref{T2}.
Therefore, to transmit message with size $e^{R}$, the required time in this case is
calculated to be 
$ \frac{2 RT}{ I(Y;A_1+A_2)_{\rm Eq. \eqref{MAC}} }+\frac{RT}{I(Y;A)_{\rm Eq. \eqref{C2}}}$.

\begin{table}[htpb]
  \caption{Secure physical layer network coding with multiple access Gaussian channel}
\label{T2}
\begin{center}
{
\renewcommand\arraystretch{1.7}
  \begin{tabular}{|c|c|c|c|} 
\hline
Time slot &Time 1 & Time 2 & Time 3 \\
\hline
Channel &$(e_1, e_2)$ &$e_3$ & $(e_4,e_5)$,$(e_6,e_7)$
\\
\hline
  \end{tabular}}
\end{center}
\end{table}

Fig. \ref{G1} gives the numerical comparison among 
$ \frac{4RT}{I(Y;A)_{\rm Eq. \eqref{C2}}}$,
$ \frac{2RT}{I(Y;A)_{\rm Eq. \eqref{C2}}}+ \frac{2RT}{I(Y;A_1,A_2)_{\rm Eq. \eqref{MAC}}}$,
$ \frac{2 RT}{2 I(Y;A_1+A_2)_{\rm Eq. \eqref{MAC}} -I(Y; A_1,A_2  )_{\rm Eq. \eqref{MAC}}}+\frac{RT}{I(Y;A)_{\rm Eq. \eqref{C2}}}$,
and
$ \frac{2 RT}{ I(Y;A_1+A_2)_{\rm Eq. \eqref{MAC}} }+\frac{RT}{I(Y;A)_{\rm Eq. \eqref{C2}}}$.
Secure network coding protocol given in Section \ref{S32} requires shorter time length for the transmission
than secure physical layer network coding protocol given in Section \ref{S33} 
in this comparison.
Since the difference is not so extensive,
secure physical layer network coding protocol given in the first paragraph of Section \ref{S33}
is useful 
when it is not easy to prepare secure shared randomness between two source nodes.
In fact, when we use the butterfly network, 
it is usual that the direct communication between two source nodes
is not easy.
In this case, such a secure shared randomness 
requires an additional cost.
However,
secure physical layer network coding protocol given in the second paragraph of Section \ref{S33}
has no advantage over the 
secure network coding protocol with MAC channel.
That is,
a simple combination of 
secure network coding and physical layer network coding
is not so useful in this case.

\begin{figure}[h]
\begin{center}
\includegraphics[scale=0.9]{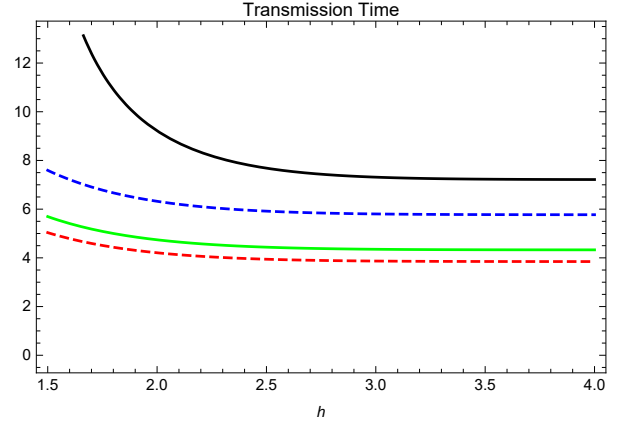}
\end{center}
\caption{Transmission Time for four schemes
when $RT=1$.
Upper solid line (Black) expresses 
the time $ \frac{2 RT}{2 I(Y;A_1+A_2)_{\rm Eq. \eqref{MAC}} -I(Y; A_1,A_2  )_{\rm Eq. \eqref{MAC}}}+\frac{RT}{I(Y;A)_{\rm Eq. \eqref{C2}}}$ of
secure physical layer network coding protocol given in the 1st paragraph of Section \ref{S33}. 
Upper dashed line (Blue) expresses 
the time $ \frac{4RT}{I(Y;A)_{\rm Eq. \eqref{C2}}}$ of
secure network coding protocol given in Section \ref{S32} without MAC channel. 
Lower dashed line (Red) expresses 
the time $ \frac{2RT}{I(Y;A)_{\rm Eq. \eqref{C2}}}+ \frac{2RT}{I(Y;A_1,A_2)_{\rm Eq. \eqref{MAC}}}$ of
secure network coding protocol given in Section \ref{S32} with MAC channel. 
Lower solid line (Green) expresses 
the time 
$ \frac{2 RT}{ I(Y;A_1+A_2)_{\rm Eq. \eqref{MAC}} }+\frac{RT}{I(Y;A)_{\rm Eq. \eqref{C2}}}$
of secure physical layer network coding protocol given in the 2nd paragraph of Section \ref{S33}. 
}
\Label{G1}
\end{figure}%


\section{Network with three sources}\label{S4}
Next, we consider the network given in Fig \ref{F1}.
This network has three source nodes $S_1$, $S_2$, and $S_3$,
three intermediate nodes $I_1$, $I_2$, and $I_3$,
and one terminal node $T$.
The aim of this network is 
secure transmission from the three source nodes to the terminal node $T$.
The source node $S_i$ intends to transmit an element $M_i \in \FF_q$ to the terminal node $T$.

\subsection{Secure network coding}\label{S4-1}
\begin{figure*}[h]
\begin{center}
\includegraphics[scale=0.5]{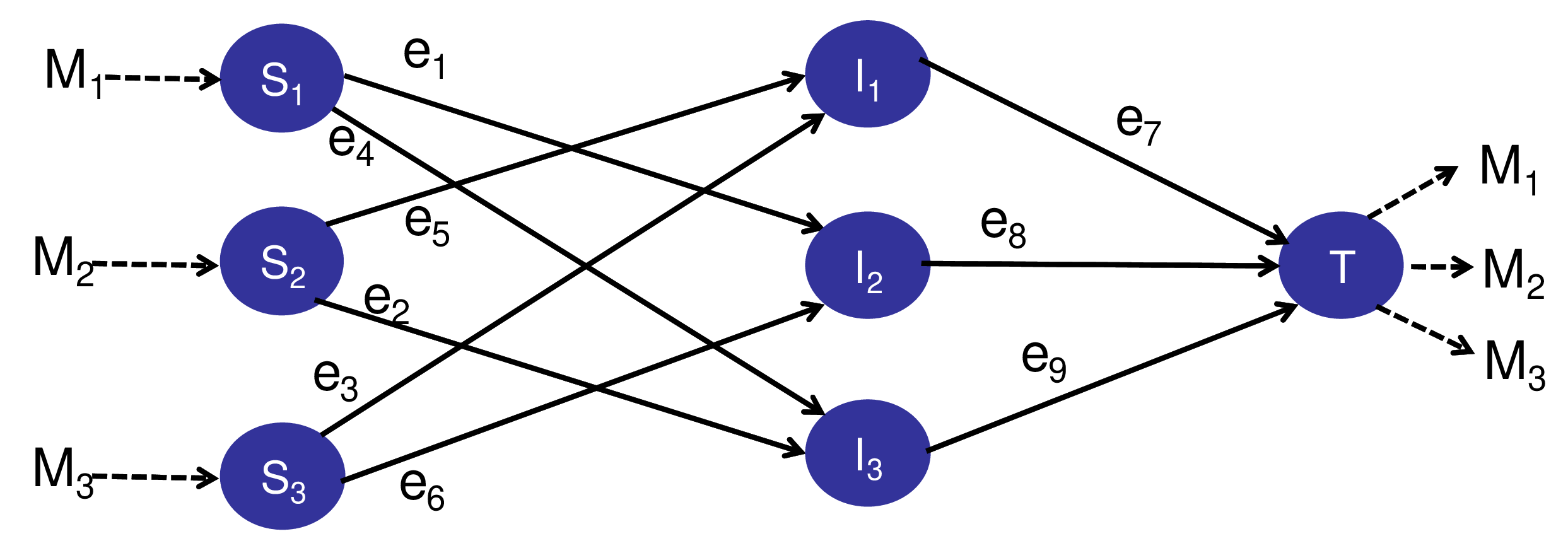}
\end{center}
\caption{Network with three sources.}
\Label{F1}
\end{figure*}%

First, we consider this network with the framework of secure network coding.
Each edge expresses a channel to transmit one element of $\FF_q$ without error.
We consider two settings.
\begin{description}
\item[(1)] Eve can eavesdrop one edge among three edges between the intermediate nodes and the
terminal node.
\item[(2)] Eve can eavesdrop one intermediate node among three intermediate nodes.
\end{description}

\subsubsection{Case (1)}\label{S4-1-1}
The following code is secure in the case (1) when $q$ is not a power of $2$.
Notice that the matrix 
$\left(
\begin{array}{ccc}
0 & 1 & 1\\
1 & 0 & 1\\
1 & 1 & 0
\end{array}
\right)$ is invertible
because 
$\left(
\begin{array}{ccc}
-1/2 & 1/2 & 1/2 \\
1/2 & -1/2 & 1/2 \\
1/2 & 1/2 & -1/2 
\end{array}
\right)$ is the inverse matrix.
Source node $S_i$ sends $M_i$ in each edge.
Each intermediate node sends the sum of the received information.
Finally, applying the inverse matrix $\left(
\begin{array}{ccc}
-1/2 & 1/2 & 1/2 \\
1/2 & -1/2 & 1/2 \\
1/2 & 1/2 & -1/2 
\end{array}
\right)$ to the received information, 
the node $T$ recovers all messages. 
In this code, each of the messages $M_1+M_2$, 
$M_2+M_3$, and $M_3+M_1$ is independent of anyone of $M_1$, $M_2$, and $M_3$.
Hence, the security in the case (1) is satisfied.
The rate of this protocol is the optimal 
even without the secrecy condition.

Next, we consider the case when $q\ge 4$ is a power of $2$.
We choose an element $e\in \FF_q$ such that $e^2+e\neq 0$,
which implies that
$\left(
\begin{array}{ccc}
 0 & 1 & 1\\
 1 & 0 & e\\
 e & e & 0
\end{array}
\right)$ is invertible
because its determinant is $ e^2+e\neq 0$.
For example, when $q=4$, since $e^2=e+1$,
the inverse matrix is
$\left(
\begin{array}{ccc}
1+e & e & e \\
1+e & e & 1 \\
e & e & 1 
\end{array}
\right)$.
Then, the following code is secure;
Source node $S_i$ sends $M_i$ in each edge.
The intermediate nodes $I_1$, $I_2$, and $I_3$
send the received information
$Z_1:= M_2+M_3 $,
$Z_2:= M_1+e M_3 $, and
$Z_3:= e M_1+e M_2 $, respectively.
Finally, applying the inverse matrix of 
$\left(
\begin{array}{ccc}
 0 & 1 & 1\\
 1 & 0 & e\\
 e & e & 0
\end{array}
\right)$
to the received information
$\left(
\begin{array}{c}
Z_1\\
Z_2\\
Z_3
\end{array}
\right)$, 
the node $T$ recovers all messages. In this code,
each of the informations $e M_1+e M_2$, 
$M_2+M_3$, and $e M_3+M_1$ is independent of anyone of $M_1$, $M_2$, and $M_3$.
Hence, the security in the case (1) is satisfied.
That is, this code guarantees the following type of security;
(T1) When the eavesdropper attacks only one of edges, 
she obtains no information for  anyone of the messages $M_1$, $M_2$ and $M_3$.

\subsubsection{Case (2)}\label{S4-1-2}
In the case (2), the following code is secure.
We use the channels between the intermediate nodes and the terminal node twice, 
but, we use the channels between the source nodes and the intermediate nodes only once.
Source node $S_i$ prepares scramble variable $L_i$.
Source node $S_i$ sends the scramble variable $L_i$ to the intermediate node 
$I_{i+1}$ via the edge $e_i$.
Source node $S_i$ sends the variable $M_i-L_i$ to Intermediate node $I_{i-1}$ via the edge $e_{3+i}$.
Here $i+1 $ and $i-1$ are regarded as elements of $\mathbb{Z}_3$.
Each intermediate node sends both received variables to the terminal node by using the channel twice.
Since the terminal node $T$ obtains information 
$L_1$, $L_2$, $L_3$, $M_1-L_1$, $M_2-L_2$, and $M_3-L_3$,
it can recover the messages $M_1=(M_1-L_1)+L_1$, $M_2=(M_2-L_2)+L_2$, and $M_3=(M_3-L_3)+L_3$.
The information on the intermediate node $I_{i}$ is the pair of $L_{i+1}$ and $M_{i-1}-L_{i-1}$, which is independent of
anyone of $M_1$, $M_2$, and $M_3$.
Hence, this code guarantees the following type of security;
(T2) When no intermediate node colludes with another node,
each intermediate node obtains no information for the messages.

\subsection{Secure physical layer network coding}\label{S4-2}
\subsubsection{Use of secure computation-and-forward}\label{S4-2-1}
Now, we assume the assumption; (T3) 
The pairs $(e_1, e_6)$, $(e_2,e_4)$, and $(e_3,e_5)$ are 
given as multiple access Gaussian channels like \eqref{MAC}.
We assume that the eavesdropper can access one of the information on the intermediate nodes,
which corresponds to Case 2 of Section \ref{S4-1}.
Then, using secure computation-and-forward, we construct a required protocol.

First, we consider the case when $q$ is not a power of $2$.
In the multiple access Gaussian channel $(e_1,e_6)$,
we employ secure computation-and-forward so that the node 
$I_2$ obtains the information $M_1+M_3$.
Similarly, 
$I_1$ and $I_3$ obtain the information $M_2+M_3$
and $M_1+M_2$, respectively.
Therefore, the information on each intermediate node
is independent of the messages $M_1$, $M_2$, and $M_3$.
In the next step,
the intermediates nodes $I_1$, $I_2$, and $I_3$ transmit their obtained information 
$M_1',M_2',$ and $M_3'$
to the terminal node $T$ via 
the multiple access Gaussian channels with three input signals.
Then, applying separate decoding, 
the terminal node $T$ recovered 
the information 
$M_1',M_2',$ and $M_3'$.
Using the method given in Section \ref{S4-1-1},
the terminal node $T$ obtains 
the original information 
$M_1,M_2,$ and $M_3$.

When $q \ge 4$ is a power of $2$,
to apply the method given in Section \ref{S4-1-1},
the node $I_2$ needs to obtain the information $M_1+e M_3$.
It can be realized by secure computation-and-forward with
a $2$-dimensional vector over the finite field $\FF_2$ 
by the prior conversion from $M_3$ to $e M_3$ in the node $S_3$
before use of the multiple access Gaussian channel $(e_1, e_6)$.
The same method is applied to
the multiple access Gaussian channels 
$(e_2,e_4)$ and $(e_3,e_5)$.
Then, the remaining part can be done in the same way as the above.

Therefore, under the framework of secure physical layer network coding,
we can realize the secure code for the attack to an intermediate node
by using secure computation-and-forward.
That is, this code guarantees the following types of security;
(T4) When no node colludes with another node and (T3) is satisfied,
each intermediate node obtains no information for the messages.
This code does not require additional random number like the code given in Section \ref{S4-1-2}.

\subsubsection{Use of computation-and-forward}\label{S4-2-2}
Next, using computation-and-forward, we construct a required protocol.
For this aim, we employ the protocol given in Section \ref{S4-1-2}.
In this protocol, at the node $T$, 
to recover $M_1$
we employ computation-and-forward of two edges $e_8$ and $e_9$.
Similarly, 
to recover $M_2$ ($M_3$),
we employ computation-and-forward of two edges $e_7$ and $e_9$ ($e_7$ and $e_8$).

\subsection{Comparison}\label{S44}
To implement these protocols as wireless communication network,
we compare the transmission speeds of 
the protocols given in Sections \ref{S4-1} and \ref{S4-2}
when each edge is given as the BPSK scheme
of 
a single-input Gaussian channel \eqref{C2},
a two-input Gaussian channel \eqref{MAC}, or
a three-input Gaussian channel \eqref{MAC}
\begin{align}
Y= h X_1+h X_2+h X_3+Z,
\label{C3}
\end{align}
where
$h \in \mathbb{C}$ are the channel fading coefficients,
$Z $ is a Gaussian complex random variable with average 0 and variance 1,
and $X_i$ is coded as $(-1)^{A_i}$ with $A_i \in \FF_2$.
In this comparison, we make the same assumptions for $h_1$, $h_2$, and $T$. 
Additionally, we assume that ideal codes given in Section \ref{S34} are available,
and that the mutual information rate triple
$(I(Y;A_1A_2A_3)_{\rm Eq. \eqref{C3}}/3,
I(Y;A_1A_2A_3)_{\rm Eq. \eqref{C3}}/3,
I(Y;A_1A_2A_3)_{\rm Eq. \eqref{C3}}/3)$ 
is available in the MAC channel \eqref{C3}
when three senders intend to send their own message to the receiver,
where
the random variables $A_1$, $A_2$, and $A_3$ are subject to the uniform distribution independently.  
Under these assumptions, we compare 
secure network coding protocol given in Section \ref{S4-1-2} 
and secure physical layer network coding protocol given in Section \ref{S4-2} 
because both protocols realize the secrecy for intermediate nods.

When we do not use multiple access Gaussian channel. 
secure network coding protocol given in Section \ref{S4-1-2} needs five time slots at least as Table \ref{T3}.
In particular, the edges $e_7$, $e_8$, and $ e_9$ need to send twice information as the remaining edges.
Therefore, to transmit message with size $e^{R}$, the required time in this case is
calculated to be $ \frac{8RT}{I(Y;A)_{\rm Eq. \eqref{C2}}}$.
When we use multiple access Gaussian channel,
secure network coding protocol given in Section \ref{S4-1-2}
can be realized with two time slots as Table \ref{T32}.
In this case, to transmit message with size $e^{R}$, the required time in this case is
calculated to be 
$ \frac{6RT}{I(Y;A_1A_2A_3)_{\rm Eq. \eqref{C3}}}+ \frac{2RT}{I(Y;A_1,A_2)_{\rm Eq. \eqref{MAC}}}$.

\begin{table}[htpb]
  \caption{Secure network coding without multiple access Gaussian channel}
\label{T3}
\begin{center}
{
\renewcommand\arraystretch{1.7}
  \begin{tabular}{|c|c|c|c|c|c|} 
\hline
Time span &Time 1 & Time 2 & Time 3  & Time 4 & Time 5   \\
\hline
Channel &$e_1, e_2, e_3$  & $e_4, e_5,e_6$ 
& {$e_7$} & $e_8$ & {$ e_9$} \\
\hline
  \end{tabular}}
\end{center}
\end{table}

\begin{table}[htpb]
  \caption{Secure network coding with multiple access Gaussian channel}
\label{T32}
\begin{center}
{
\renewcommand\arraystretch{1.7}
  \begin{tabular}{|c|c|c|} 
\hline
Time span &Time 1 & Time 2   \\
\hline
Channel &$(e_1, e_6)$, $(e_2,e_4)$,$( e_3,e_5)$ 
& {$(e_7,e_8, e_9)$} \\
\hline
  \end{tabular}}
\end{center}
\end{table}

Secure physical layer network coding protocol given in Section \ref{S4-2-1} 
can be realized only with two time slots as in Table \ref{T4},
where the pairs $(e_1, e_2)$, $(e_4,e_5)$, and $(e_6,e_7)$
are realized by secure computation-and-forward
based on the Gaussian MAC channel \eqref{MAC}.
Therefore, to transmit message with size $e^{R}$, the required time in this case is
calculated to be 
$ \frac{3RT}{I(Y;A_1A_2A_3)_{\rm Eq. \eqref{C3}}}+ 
 \frac{RT}{2 I(Y;A_1+A_2)_{\rm Eq. \eqref{MAC}} -I(Y; A_1,A_2  )_{\rm Eq. \eqref{MAC}}}$.

\begin{table}[htpb]
  \caption{Secure physical layer network coding with secure computation-and-forward}
\label{T4}
\begin{center}
{
\renewcommand\arraystretch{1.7}
  \begin{tabular}{|c|c|c|} 
\hline
Time span &Time 1 & Time 2  \\
\hline
Channel &$(e_1, e_6)$, $(e_2,e_4)$, $(e_3,e_5)$ 
&$(e_7, e_8, e_9)$
\\
\hline
  \end{tabular}}
\end{center}
\end{table}

Another secure physical layer network coding protocol given in Section \ref{S4-2-2} 
can be realized only with two time slots as in Table \ref{T5},
where the pairs $(e_1, e_2)$, $(e_4,e_5)$, and $(e_6,e_7)$
are realized by secure computation-and-forward
based on the Gaussian MAC channel \eqref{MAC}.
Therefore, to transmit message with size $e^{R}$, the required time in this case is
calculated to be 
$ \frac{3RT}{I(Y;A_1+A_2)_{\rm Eq. \eqref{MAC}}}+ \frac{2RT}{I(Y;A_1,A_2)_{\rm Eq. \eqref{MAC}}}$.

\begin{table}[htpb]
  \caption{Secure physical layer network coding with computation-and-forward}
\label{T5}
\begin{center}
{
\renewcommand\arraystretch{1.7}
  \begin{tabular}{|c|c|c|c|c|} 
\hline
Time span &Time 1 & Time 2& Time 3& Time 4   \\
\hline
Channel &$(e_1, e_6)$, $(e_2,e_4)$,$( e_3,e_5)$ 
& $(e_8, e_9)$ & $(e_7, e_9)$ & $(e_7, e_8)$ 
\\
\hline
  \end{tabular}}
\end{center}
\end{table}

Fig. \ref{G2} gives the numerical comparison among 
$ \frac{3RT}{I(Y;A_1A_2A_3)_{\rm Eq. \eqref{C3}}}+ \frac{RT}{2 I(Y;A_1+A_2)_{\rm Eq. \eqref{MAC}} -I(Y; A_1,A_2  )_{\rm Eq. \eqref{MAC}}}$,
$ \frac{3RT}{I(Y;A_1+A_2)_{\rm Eq. \eqref{MAC}}}+ \frac{2RT}{I(Y;A_1,A_2)_{\rm Eq. \eqref{MAC}}}$,
$ \frac{8RT}{I(Y;A)_{\rm Eq. \eqref{C2}}}$,
and
$ \frac{6RT}{I(Y;A_1A_2A_3)_{\rm Eq. \eqref{C3}}}+ \frac{2RT}{I(Y;A_1,A_2)_{\rm Eq. \eqref{MAC}}}$.
Codes of secure physical layer network coding protocol given in Section \ref{S4-2} 
require shorter time length for the transmission than 
secure network coding protocol given in Section \ref{S4-1-2} 
in this comparison
when the coefficient $h$ is larger than about $1.7$.
This comparison shows the advantage of the secure physical layer network coding protocol given in Section \ref{S4-2-1} 
over the secure network coding protocol given in Section \ref{S4-1-2}.
Also, this comparison indicates the advantage of the simple combination of 
secure network coding and physical layer network coding given in Section \ref{S4-2-2} 
over the secure network coding protocol given in Section \ref{S4-1-2} with MAC channel.

\begin{figure}[h]
\begin{center}
\includegraphics[scale=0.9]{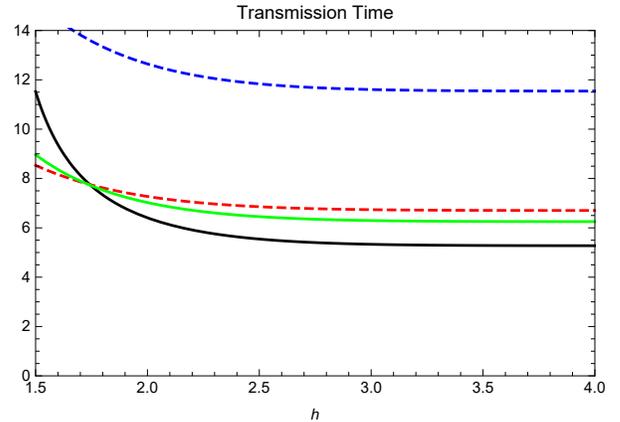}
\end{center}
\caption{Transmission Time for four schemes
when $RT=1$.
Solid line (Black) expresses 
the time $ \frac{3RT}{I(Y;A_1A_2A_3)_{\rm Eq. \eqref{C3}}}+ 
 \frac{RT}{2 I(Y;A_1+A_2)_{\rm Eq. \eqref{MAC}} -I(Y; A_1,A_2  )_{\rm Eq. \eqref{MAC}}}$
 of
secure physical layer network coding protocol given in Section \ref{S4-2-1}. 
Solid line (Green) expresses 
the time 
$ \frac{3RT}{I(Y;A_1+A_2)_{\rm Eq. \eqref{MAC}}}+ \frac{2RT}{I(Y;A_1,A_2)_{\rm Eq. \eqref{MAC}}}$
of secure physical layer network coding protocol given in Section \ref{S4-2-2}. 
Upper dashed line (Blue) expresses 
the time $ \frac{8RT}{I(Y;A)_{\rm Eq. \eqref{C2}}}$ of
secure network coding protocol given in Section \ref{S4-1-2} without MAC channel. 
Lower dashed line (Red) expresses 
the time $ \frac{6RT}{I(Y;A_1A_2A_3)_{\rm Eq. \eqref{C3}}}+ \frac{2RT}{I(Y;A_1,A_2)_{\rm Eq. \eqref{MAC}}}$ of
secure network coding protocol given in Section \ref{S4-1-2} with MAC channel.
Solid line (Black), Solid line (Green), and Lower dashed line (Red) 
intersect around$h=1.7$.
}
\Label{G2}
\end{figure}%

\section{Conclusion}
We have discussed the advantages of 
secure physical layer network coding over secure network coding.
To clarify this advantage, we have addressed two typical networks.
One is the butterfly network (Fig \ref{F8}), and 
the other is a network with three source nodes (Fig. \ref{F1}).
That is, we have given a concrete protocol that efficiently works 
on these examples. 
We have also compared transmission times of proposed codes.

In these examples, secure physical layer network coding can realize the secrecy against intermediate nodes.
Therefore, we can consider that secure physical layer network coding is useful 
when we realize the secrecy against intermediate nodes.
In fact, when adversary attacks an intermediate node 
secure network coding requires more randomness than 
when adversary attacks an edge.
Further, there are still a small number of applications of secure physical layer network coding.
Hence, it is a future study to find more fruitful applications of secure physical layer network coding.

\section*{Acknowledgments}
The author is grateful to 
Prof. \'{A}ngeles Vazquez-Castro,
Prof. Tadashi Wadayama, and  
Dr. Satoshi Takabe for discussions on secure physical layer network coding. 
The author is also grateful to 
Dr. Go Kato and Prof. Masaki Owari
for discussions on secure network coding. 
The work reported here was supported in part by 
the JSPS Grant-in-Aid for Scientific Research 
(A) No.17H01280, (B) No. 16KT0017, (C) No. 16K00014, 
and Kayamori Foundation of Informational Science Advancement.

\end{document}